\documentclass[twocolumn,tighten]{aastex631}

\pdfoutput=1 %for arXiv submission
\usepackage{amsmath,amstext,longtable}
\usepackage[T1]{fontenc}
\usepackage{apjfonts} 

\usepackage[figure,figure*]{hypcap}
\received{TBD}
\revised{TBD}
\accepted{TBD}
\submitjournal{ApJ}

\shorttitle{Mass Function Simulations}
\shortauthors{Raghu et al.}

\usepackage{color,amsmath,longtable}

\usepackage{CJK}

\begin{document}
\begin{CJK*}{UTF8}{gbsn}

\title{Simulating Brown Dwarf Observations for Various Mass Functions, Birthrates, and Low-mass Cutoffs}

\correspondingauthor{Yadukrishna Raghu}
\email{yadukrishnaraghu01@gmail.com}

\author[0000-0001-9778-7054]{Yadukrishna Raghu}
\affiliation{IPAC, Mail Code 100-22, California Institute of Technology, 1200 E. California Blvd., Pasadena, CA 91125, USA}
\affiliation{Washington High School, 38442 Fremont Blvd, Fremont, CA 94536, USA}
\affiliation{Backyard Worlds: Planet 9}

\author[0000-0003-4269-260X]{J.\ Davy Kirkpatrick}
\affiliation{IPAC, Mail Code 100-22, California Institute of Technology, 1200 E. California Blvd., Pasadena, CA 91125, USA}
\affiliation{Backyard Worlds: Planet 9}

\author[0000-0001-7519-1700]{Federico Marocco}
\affiliation{IPAC, Mail Code 100-22, California Institute of Technology, 1200 E. California Blvd., Pasadena, CA 91125, USA}
\affiliation{Backyard Worlds: Planet 9}

\author[0000-0001-5072-4574]{Christopher R.\ Gelino}
\affiliation{NASA Exoplanet Science Institute, Mail Code 100-22, California Institute of Technology, 770 S. Wilson Avenue, Pasadena, CA 91125, USA}

\author[0000-0001-8170-7072]{Daniella C.\ Bardalez Gagliuffi}
\affiliation{Department of Astrophysics, American Museum of Natural History, Central Park West at 79th Street, New York, NY 10024, USA}
\affiliation{Backyard Worlds: Planet 9}

\author[0000-0001-6251-0573]{Jacqueline K.\ Faherty}
\affiliation{Department of Astrophysics, American Museum of Natural History, Central Park West at 79th Street, New York, NY 10024, USA}
\affiliation{Backyard Worlds: Planet 9}

\author[0000-0003-1785-5550]{Steven D.\ Schurr}
\affiliation{IPAC, Mail Code 100-22, California Institute of Technology, 1200 E. California Blvd., Pasadena, CA 91125, USA}

\author[0000-0002-6294-5937]{Adam C.\ Schneider}
\affiliation{United States Naval Observatory, Flagstaff Station, 10391 West Naval Observatory Road, Flagstaff, AZ 86005, USA}
%\affiliation{Department of Physics and Astronomy, George Mason University, MS3F3, 4400 University Drive, Fairfax, VA 22030, USA}
\affiliation{Backyard Worlds: Planet 9}

\author[0000-0002-1125-7384]{Aaron M. Meisner}
\affiliation{NSF's National Optical-Infrared Astronomy Research Laboratory, 950 N. Cherry Ave., Tucson, AZ 85719, USA}
\affiliation{Backyard Worlds: Planet 9}

\author[0000-0002-2387-5489]{Marc J.\ Kuchner}
\affiliation{NASA Goddard Space Flight Center, Exoplanets and Stellar Astrophysics Laboratory, Code 667, Greenbelt, MD 20771, USA}
\affiliation{Backyard Worlds: Planet 9}

\author[0000-0002-5253-0383]{Hunter Brooks}
\affiliation{Department of Astronomy and Planetary Science, Northern Arizona University, Flagstaff, AZ 86011, USA}
\affiliation{Backyard Worlds: Planet 9}

\author[0000-0002-2466-865X]{Jake Grigorian}
\affiliation{University of Southern California, University Park Campus, Los Angeles, CA 90089, USA}
\affiliation{Saint Francis High School, 200 Foothill Blvd., La Ca\~nada, CA 91011, USA}

%\authos[0000-0002-2592-9612]{Jonathon Gagne}
%\author[0000-0002-6523-9536]{Adam Burgasser}

%\author[0000-0003-0580-7244]{Katelyn Allers}
%\author[0000-0001-7896-5791]{Dan Caselden}
%\author[0000-0003-2478-0120]{Sarah Casewell}
\author{The Backyard Worlds: Planet 9 Collaboration}

\begin{abstract}
  
After decades of brown dwarf discovery and follow-up, we can now infer the functional form of the mass distribution within 20 parsecs, which serves as a constraint on star formation theory at the lowest masses. Unlike objects on the main sequence that have a clear luminosity-to-mass correlation, brown dwarfs lack a correlation between an observable parameter (luminosity, spectral type, or color) and mass. A measurement of the brown dwarf mass function must therefore be procured through proxy measurements and theoretical models.  We utilize various assumed forms of the mass function, together with a variety of birthrate functions, low-mass cutoffs, and theoretical evolutionary models, to build predicted forms of the effective temperature distribution. We then determine the best fit of the observed effective temperature distribution to these predictions, which in turn reveals the most likely mass function. We find that a simple power law ($dN/dM \propto M^{-\alpha}$) with $\alpha \approx 0.5$ is optimal. Additionally, we conclude that the low-mass cutoff for star formation is $\lesssim0.005M_{\odot}$. We corroborate the findings of \cite{burgasser2004} which state that the birthrate has a far lesser impact than the mass function on the form of the temperature distribution, but we note that our alternate birthrates tend to favor slightly smaller values of $\alpha$ than the constant birthrate. Our code for simulating these distributions is publicly available. As another use case for this code, we present findings on the width and location of the subdwarf temperature gap by simulating distributions of very old (8-10 Gyr) brown dwarfs.
\end{abstract}


\section{Introduction}

First detected in 1995 (\citealt{oppenheimer1995}; \citealt{nakajima1995}), brown dwarfs, defined to be objects below the Hydrogen-1 fusing limit of $\sim0.075 \text{ M}_{\odot}$ (\citealt{kumar1963, hayashi1963}), bridge the mass gap between hydrogen-fusing stars and exoplanets. Despite the substantial advancements of understanding that the field of brown dwarf astronomy have experienced in the past two decades through infrared missions such as the National Aeronautics and Space Administration's (NASA) Wide-field Infrared Survey Explorer (hereafter, WISE; \citealt{wright2010}) and NASA's Spitzer Space Telescope (\citealt{werner2004}), there is still an abundance of open questions regarding many aspects of brown dwarfs. Examples are the exact formation mechanisms that prevail in different mass regimes, as well as the low-mass cutoff of this formation process. Answering these questions and improving the theory necessitates additional brown dwarf observational data, be it spectroscopic or photometric. However, observing brown dwarfs is an ordeal in itself, with some known examples as faint as $\sim28$ mag at 1.15 microns (James Webb Space Telescope F115W filter)  and an estimated distance greater than $570$ pc (\citealt{nonino2023}). However, the distance itself is not necessarily the defining factor in a brown dwarf's faintness, since exceedingly close brown dwarfs have also been observed to be especially faint. The leading example is WISE J085510.83$-$071442.5, which is confidently estimated to have a $J_{MKO}>24.0$ mag (\citealt{faherty2014}) at $\sim2.3$ pc (\citealt{kirkpatrick2021}).
Compounding this dilemma is the considerable challenge of reliably measuring the physical properties of the object, such as mass, age, and temperature. Those familiar with the methods of stellar astronomy will recall that the masses of main-sequence stars can be derived with little uncertainty using only a few common observables such as color, absolute magnitude, or spectral type. However, brown dwarfs do not possess such simple relations between physically observable quantities and mass. A brown dwarf of a certain temperature or spectral type may have a range of possible masses. Such a coupling of parameters is a consequence of cooling over astronomical timescales, as brown dwarfs cool continually throughout their lifetimes. This means a massive old brown dwarf that has cooled can have a similar temperature to a lower-mass young brown dwarf. 
\par 
Despite this observational barrier, techniques have been formulated to directly derive the mass of a brown dwarf, yet can often only be employed in rare, opportune cases. 
\par For resolvable brown dwarf binary systems, one may leverage the orbital dynamics of the system and then solve for the mass of the brown dwarf. To measure the mass of our desired object, $M_2$, we need the semi-major axes of both orbits, $a_1$ and $a_2$, as well as the orbital period, $P$, inclination $i$, and the gravitational constant $G$ (\citealt{bob}).
\begin{equation}
    M_2=\frac{4\pi^2(a_1+a_2)^3}{GP^2\left(\frac{a_2}{a_1}+1\right)\cos^3i}
   \label{eqn:dirmass}
\end{equation}

\par Alternatively, there is microlensing, in which we observe a brown dwarf passing between a background light source and the observer. Since the mass of the transiting brown dwarf affects the path of the light emitted by the background star due to gravitational lensing, one can determine the mass of the lens by measuring the displacement and amplification of the light emitted by the background object (\citealt{DominikSahu2000}, \citealt{cushing2014}).
\par
Nevertheless, occasions in which we are able to apply these methods are exceptionally rare. The current methods of direct observation would provide only a handful of directly observed or inferred masses in any volume-limited sample. One notable exception to this are brown dwarf constituents of a young star formation region or moving group, for which a robust age can be assumed for all objects. Drawbacks in this case are a higher reliance on evolutionary models at young ages, interstellar reddening (since these clusters are more distant than the local sample and are often still enshrouded in dust), and difficulties in resolving close multiple systems and assuring completeness of the brown dwarf sample.
\par 
Since we find ourselves at an impasse when pursuing direct paths of constructing and validating the mass distributions for brown dwarfs, we instead choose to use the temperature of brown dwarfs as a proxy measurement to indirectly constrain the brown dwarf mass function. Although accessing temperature data is not so simple for brown dwarfs as for main-sequence stars, it is a far more accessible measurement than direct brown dwarf mass measurements.We compare our predicted distributions with the observational distribution of brown dwarf temperatures (e.g., \citealt{kirkpatrick2019}). In our study, we construct theoretical temperature distributions with the inverse transform method, assuming theoretical mass and age distributions that have found success in previous literature (\citealt{kirkpatrick2019, kirkpatrick2021, johnson2021}). 

Then, we make use of a variety of brown dwarf evolutionary models, all with somewhat different assumed internal physics, to calculate the temperature of each object. Combining these calculated temperatures across a simulated population allows us to build its temperature distribution. 
From here, the problem becomes one of optimization, as we seek to obtain the particular set of parameters (functional form of the mass function, birthrate, low-mass cutoff, and evolutionary model suite) that results in a temperature distribution whose shape most accurately fits the observed distribution. Our extensive sampling of the permutations of parameters constrains the functional form of the brown dwarf mass function.

\par 
 In \S 2 we present our chosen mass functions, which we need for the implementation of the inverse transform methodology. We also discuss the topic of the low-mass cutoff for brown dwarfs. In \S 3 we explore different proposed age distributions, and \S 4 examines the evolutionary models we use as well as their physical implications. \S 5 combines the tools developed in the three preceding sections to create our simulated brown dwarf populations and their temperature distributions. \S 6 contains a comparison of our simulated stellar populations to our empirical data, as well as an analysis of the impact of certain parameters on the derived temperature distribution. In \S 7 we provide our concluding remarks and possible avenues of future research.

\section{Mass Distributions}
Past studies on the functional form of the stellar mass distribution are replete with power law formalisms.
Power laws have been found to represent the functional forms seen in the 0.3-10 $M_\odot$ (\citealt{salpeter1955}) and 0.1-63 $M_\odot$ (\citealt{millerscalo1979}) regimes. 
In addition to the two power law mass functions needed to describe (higher mass) stars, there may be a third, separate power law form needed for the (lower mass) brown dwarf regime (\citealt{kirkpatrick2021}), along with a fourth in the M dwarf regime (\citealt{kirkpatrick2023}). The youthfulness of the field of brown dwarf science combined with a lack of ample data sets has meant that many functional forms have been theorized. Some examples of these are the log-normal (\citealt{chabrier2023, chabrier2003a, chabrier2003b, chabrier2001}), and the bi-partite power law from \cite{kroupa2013}, their equation 55. The physics of the brown dwarf formation mechanism(s) will ultimately determine the way that the mass in the natal cloud is distributed among the birthed objects. 
Each birth mechanism results in a different mass distribution for brown dwarfs; for example, a power-law arises from stellar birth physics that is independent of the size of the natal cloud (\citealt{kirkpatrick2021}). On the other hand, the log-normal implies a set of multiplicative birth parameters (\citealt{kapteyn1903}).
\par

\par As previous investigations of the substellar mass function have found simple power laws to be the favored functional form (\citealt{kirkpatrick2019, kirkpatrick2021}), we also choose to adopt a simple power law as our proposed mass distribution, or Probability Distribution Function (PDF), of brown dwarfs. The functional form of this simple power law is written as follows, in Equation \ref{eqn:PDF}, with parameter $\alpha$, constant of normalization $C_N$, and input mass $\mathcal{M}$.
\begin{equation}
    \text{PDF}(\mathcal{M})=C_N\mathcal{M}^{-\alpha}
    \label{eqn:PDF}
\end{equation}

\subsection{Low-Mass Cutoff}
A crucial parameter of brown dwarf formation is the value of the low-mass cutoff, which has been shown to be no higher than $\sim10M_{Jup}$ (\citealt{kirkpatrick2021}). Objects such as WISE J085510.83$-$071442.5, which is estimated to a have a mass between $1.5M_{Jup}$ and $8M_{Jup}$ depending on its age (\citealt{leggett2017}), along with objects identified in young moving groups (see below), almost certainly push this limit lower, as seen by derived low-mass cutoffs of $\sim4M_{Jup}$ in \cite{bate2005}.

The lowest mass at which brown dwarfs form is a  fundamental property of star formation at the very edge of our theoretical understanding of brown dwarfs. Notably, a lower mass cutoff not only extends the mass range in which brown dwarfs may form, but also shifts the mode of the distribution to lower masses. 
These faint objects that would populate the low mass end of the mass function are predominantly late-T and Y dwarfs -- as seen in Figure 6 of \cite{burgasser2004}. 
 The observed space density in temperature bins below $750 K$ has the greatest deciding influence on the value of the low-mass cutoff, as lower-mass cutoffs will more heavily populate objects at the lowest temperatures. Thus, data at these coolest temperatures will be most influential in determining the low-mass cutoff.

 Due to the faint nature of late-T and Y dwarfs it is difficult to complete a volume-limited sample with sufficient statistics to provide a robust space density measurement. Since the lowest temperature bins are of paramount importance for the evaluation of the low-mass cutoff, we therefore need additional discoveries of faint, cold Y dwarfs in order to further constrain the value of the low-mass cutoff. 
 
 For this study, we choose $0.01 \text{ M}_{\odot}$, $0.005 \text{ M}_{\odot}$, and $0.001 \text{ M}_{\odot}$ as the low-mass cutoffs within our framework, as was done in \cite{kirkpatrick2021}. These  values produce populations that include low-mass brown dwarfs that either straddle or are below the deuterium-burning limit ($\sim13M_{Jup}$; \citealt{spiegel2011}). 
 There are precedents for such brown dwarfs. Take, for example, the low-mass brown dwarfs SIMP J013656.5+093347.3 and 2MASSW J2244316+204343. 
 SIMP J013656.5+093347.3 is a young early-T dwarf with an estimated mass of $12.7 \pm 1.0M_{Jup}$, derived
 using its moving group association and evolutionary models (\citealt{saumonmarley2008}), and a trigonometric distance of $6.139 \pm 0.037  \text{ pc}$ (\citealt{gagne2017}). 2MASSW J2244316+204343 is a mid-L dwarf with a mass of $10.46 \pm 1.49M_{Jup}$ (\citealt{faherty2016}), also derived from evolutionary models, and a kinematic distance of $18.5\pm 1.2\text{ pc}$ (\citealt{liu2016}). However, both objects are close to the Solar System and therefore less of a challenge for the current instrumentation to observe, unlike further, fainter brown dwarfs.
\par

 \subsection{Deriving the Inverse CDF}

Integrating Equation \ref{eqn:PDF}, supposing $\alpha\neq 1$, to find our Cumulative Distribution Function (CDF), we get the following expression, where $M$ is the mass parameter. Our CDF, once normalized and inverted, will serve as a key component of the inverse transform method which we utilize. 
Here, $m_2$ is the high mass cutoff, defined to be $0.1 M_{\odot}$, and $m_1$ we vary between $0.01 M_{\odot}$, $0.005 M_{\odot}$, $0.001 M_{\odot}$.

\begin{equation}
    \text{CDF}(M)=C_N\int^M_{m_1}\mathcal{M}^{-\alpha}d\mathcal{M}=\frac{C_N(M^{1-\alpha}-m_1^{1-\alpha})}{1-\alpha}
\end{equation}
 In order to derive our constant of normalization, $C_N$, we need our CDF to evaluate to $1$ given the higher mass limit.  Solving for the constant, we find the following:
 \begin{equation}
    \text{CDF}(M=m2)=1=\frac{C_N(m_2^{1-\alpha}-m_1^{1-\alpha})}{(1-\alpha)}
\end{equation}
\begin{equation}
    C_N=\frac{1-\alpha}{(m_2^{1-\alpha}-m_1^{1-\alpha})}
\end{equation}
 
Similarly, when $\alpha=1$ in Equation \ref{eqn:PDF}, the constant of normalization is the following.

\begin{equation}
    C_N=\frac{1}{\ln(m_2)-\ln(m_1)}
\end{equation}

\begin{figure*}[ht!]
\centering
\includegraphics[scale=0.75]{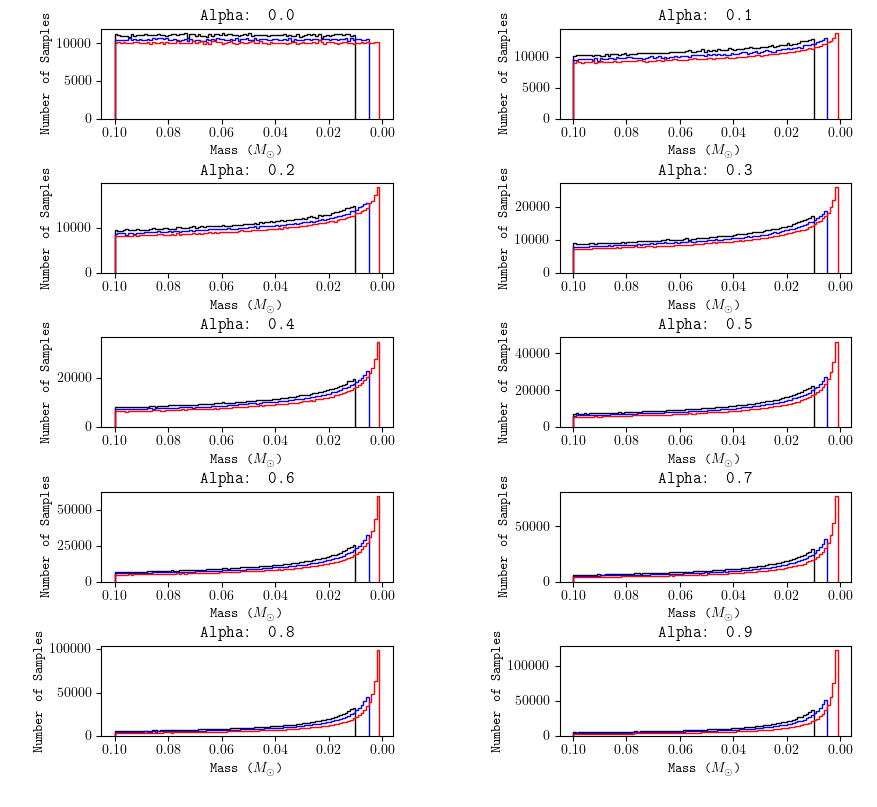}
\caption{Sampled histograms of the $\text{CDF}^{-1}$, with the $\alpha$ value ranging from $0$ to $0.9$, in increments of $0.1$. Red is the $0.001\text{ M}_{\odot}$ low-mass cutoff, blue is the $0.005\text{ M}_{\odot}$ low-mass cutoff, and black is the $0.01\text{ M}_{\odot}$ low-mass cutoff. \label{fig:mass_inverse_CDF}}
\end{figure*}

Once inverted and with the value of $C_N$ inserted, the equation for the $\text{CDF}^{-1}$ becomes the following (\citealt{kirkpatrick2019}).  
\begin{equation}
    \text{CDF}^{-1}(x) = \begin{cases} \displaystyle[x(m_2^{1-\alpha}-m_1^{1-\alpha})+m_1^{1-\alpha}]^{\frac{1}{1-\alpha}}, \text{ for } \alpha \neq 1 \nonumber\\\\ \displaystyle e^{x[ln(m_2)-ln(m_1)]+ln(m1)}, \text{ for } \alpha = 1. \nonumber \end{cases}
\end{equation}
Here, $x\in\mathcal{U}_{[0,1]}$, meaning $x$ is randomly sampled from the uniform distribution between 0 and 1. Histograms of the $\text{CDF}^{-1}$ sampled $10^6$ times per low-mass cutoff threshold with various $\alpha$ values are shown in Figure \ref{fig:mass_inverse_CDF}.

\section{Age Distributions\label{sec:age_distributions}}

\begin{figure*}[ht!]
\centering
\includegraphics[scale=0.5]{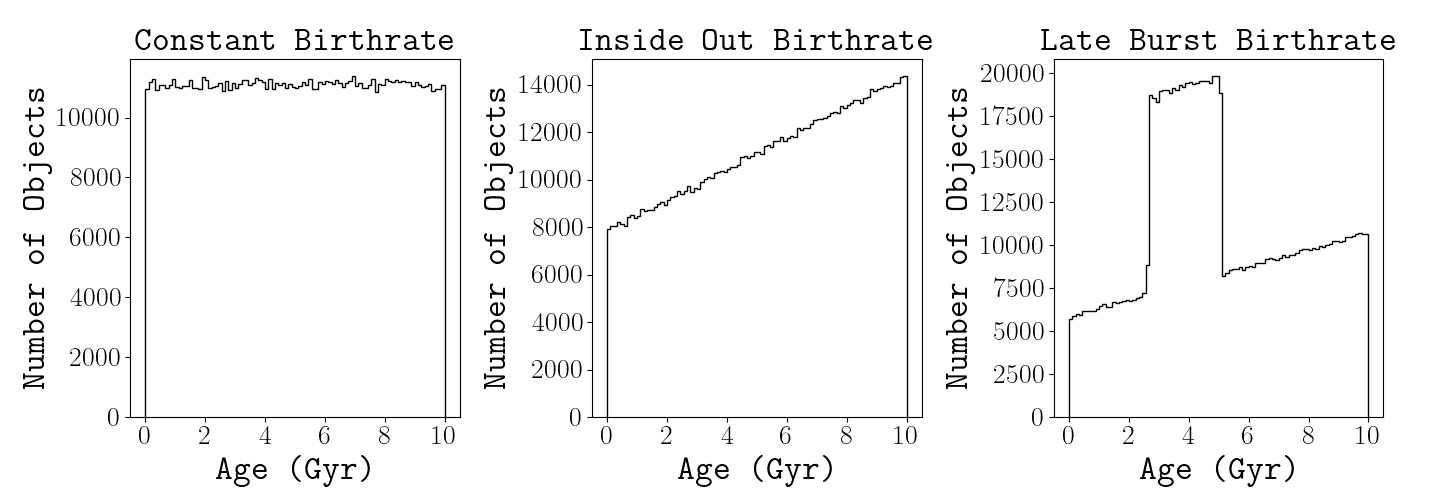}
\caption{Histograms with $10^6$ samples of the $\text{CDF}^{-1}$ for each of our three different birthrates. Note that these graphs display the birthrates after having switched from the time domain to the age domain.\label{fig:birthrates}}
\end{figure*}

We employ three different potential birthrate distributions in our main analysis (Figure~\ref{fig:birthrates}). In \S 3.1-3.3 we state the functional form of each birth time distribution and provide a few remarks on their underlying physics. Our study considers the last 10 Gyr out of the 15 Gyr of Galactic Disk stellar formation activity modeled in \cite{johnson2021}, from which comes our Inside-Out and Late-Burst birthrates. Since the evolutionary models we use (see \S \ref{sec:models}) depend on age as opposed to time we convert each time distribution into an age distribution by the following coordinate transformation, in which $\mathcal{A}$ and $\mathcal{T}$ are the age and time parameters, respectively, all in units of Gyr. 
\begin{equation}
    \text{PDF}(\mathcal{A})=\text{PDF}(15-\mathcal{T})
    \label{eqn:coordtrans}
\end{equation}
We calculate the normalized $\text{CDF}^{-1}$ of each of our proposed age distributions for later use in \S 4. Detailed studies on stellar formation processes and history can be found in (\citealt{johnson2021}). 
However, for our purposes we use age distributions only as an auxiliary measurement in our study of mass distributions, especially since ultimately the age distribution of a brown dwarf population has an undersized influence on its temperature distribution (\citealt{burgasser2004}). 

\subsection{Constant Distribution}
The constant birthrate function is a common starting point by virtue of its inherent simplicity. A constant distribution implies that the Galaxy's star formation processes have been consistently efficient and have had sufficient star forming material from its nascence to present-day. We adopt the following functional form for the constant distribution, in which $C_\mathcal{T}$ is the eponymous constant of star formation.
\begin{equation}
    \text{PDF}(\mathcal{T})\propto C_\mathcal{T} 
\end{equation}
 The value of $C_\mathcal{T}$ is not of much significance in our study as we ultimately normalize our CDF.
 \par
 Given that the constant age distribution is, as its name implies, constant, it does not depend on any time parameter $\mathcal{T}$. Therefore in order to convert it from an age distribution we simply change $C_\mathcal{T}$ to $C_\mathcal{A}$ to indicate the change from a constant of time to a constant of age, instead of executing the coordinate transformation outlined in Equation \ref{eqn:coordtrans}. 
 \begin{equation}
    \text{PDF}_C(\mathcal{A})\propto C_\mathcal{A}   
\end{equation}
\par The $\text{CDF}^{-1}$ for our constant distribution can be attained in a manner similar to \S 2.2. 

%\begin{equation}
%    \text{CDF}(A)=C_NC_\mathcal{A}\int^A_{a_1}d\mathcal{A}=C_NC_\mathcal{A}(A-a_1)
%\end{equation}

%\begin{equation}
%    \text{CDF}(A=a2)=1=C_NC_\mathcal{A}(a_2-a_1)
%\end{equation}

\begin{equation}
    \text{CDF}^{-1}_C(x)=x(a_2-a_1)+a_1
\end{equation}
We choose to leave $CDF^{-1}_{C}$ in terms of $a_1$ to $a_2$, the minimum and maximum ages in Gyr respectively, as our study explores more than one age range, see Appendix A.
\subsection{Inside-Out Distribution}

The Inside-Out age distribution represents a sample population where star formation initiates within the central regions of the Galaxy and propagates outward with time (\citealt{bird2013}). 
The functional form of the distribution is the following (\citealt{johnson2021}). 
\begin{equation}
    \text{PDF}(\mathcal{T})\propto\left(e^{\frac{-t}{15}}(1-e^{\frac{-t}{2}})\right)
    \label{eqn:insouttrue}
\end{equation}

We choose to linearly approximate this time distribution using Equation \ref{eqn:insouttime}, since inverting the CDF of Equation \ref{eqn:insouttrue} requires the use of the computationally expensive error function $\left(\text{erf}(x)=\frac{2}{\sqrt{\pi}}\int^x_0e^{-t^2}dt\right)$. Moreover, the original functional form is already sufficiently linear between our age bounds of 0 to 10 Gyr such that our approximation retains much of the original shape of the function. 
\begin{equation}
    \text{PDF}_{IO}(\mathcal{T})\propto(-0.03\mathcal{T}+0.81)
    \label{eqn:insouttime}
\end{equation}
By using Equation \ref{eqn:coordtrans} to convert this time distribution to an age distribution, we arrive at the following functional forms:
\begin{equation}
    \text{PDF}_{IO}(\mathcal{A})\propto(0.03\mathcal{A}+0.36)
    \label{eqn:insout}
\end{equation}
Integrating and normalizing this PDF yields the following CDF for the Inside-Out birthrate.
%\begin{equation}
%    \text{CDF}_{IO}(A)=\int^A_{0}\text{PDF}_{IO}(\mathcal{A})d\mathcal{A}=C_N\left(\frac{0.03}{2}A^2+0.36A\right)
%\end{equation}
%\begin{equation}
%    \text{CDF}_{IO}\left(a_2=10\right)=1=5.1C_N
%\end{equation}
\begin{equation}
    \text{CDF}_{IO}(A)=\frac{1}{5.1}\left(\frac{0.03}{2}A^2+0.36A\right)=x
\end{equation} 
We solve for $A$ as we have done previously to derive the inverse form of the CDF.
The negative branch of the solution is disregarded as a negative age is physically inconceivable. 

\begin{equation}
    \text{CDF}^{-1}_{IO}(x)=-12+\sqrt{(144+340x)}
\end{equation}

\subsection{Late-Burst Distribution}
Galactic star formation need not have followed a constant rate, or even one that varies linearly like the Inside-Out. In the past 10 Gyr it is possible that periods of the star formation history of our Galaxy have been more intense than others, with otherwise linear reduction of the birthrate, as seen in the Inside-Out. This manifests itself as bursts of increased star formation, possibly due to gravitational perturbances from the Sagittarius dwarf galaxy (\citealt{ruiz2020}) or from an earlier galactic merger incident inducing a starburst on our Galaxy (\citealt{helmi2020}). 

\par
The Late-Burst model accounts for such a period of starburst with a spike in the disk's total birthrate between ages of 2.65 and 5.10 Gyr. Equation \ref{eqn:latebursttrue} displays the mathematical expression for the Late-Burst model. For ease of inversion, we approximate the Late-Burst as Equation \ref{eqn:lateburst}, defined in terms of the two previous birthrates, $PDF_C$ and $PDF_{IO}$.  
\begin{equation}
    \text{PDF}(\mathcal{T})\propto\left(e^{\frac{-\mathcal{T}}{15}}(1-e^{\frac{\mathcal{T}}{2}}\right)\left(1+1.5e^{-\frac{(\mathcal{T}-11.2)^2}{2}}\right)
    \label{eqn:latebursttrue}
\end{equation}

\begin{equation}
    \text{PDF}_{LB}(A) \propto \begin{cases} \displaystyle PDF_{IO}, \text{ for 0-10 Gyr}  \\\\ \displaystyle PDF_{IO}+PDF_{C}, \text{ for 2.65-5.10 Gyr}
    \end{cases}
    \label{eqn:lateburst}
\end{equation}

 In order to extract a meaningful $CDF^{-1}$ from this, we integrate each piece of the Late-Burst and allocate samples to preserve the ratio between the two pieces. Thus, the $CDF^{-1}_{LB}$ is a mixture of $CDF^{-1}_{IO}$ and $CDF^{-1}_{C}$ whose individual sample sizes depend on the area under each individual PDF.

\section{Evolutionary Models\label{sec:models}}

\begin{figure*}[ht!]
    \centering
    \includegraphics[scale=0.6]{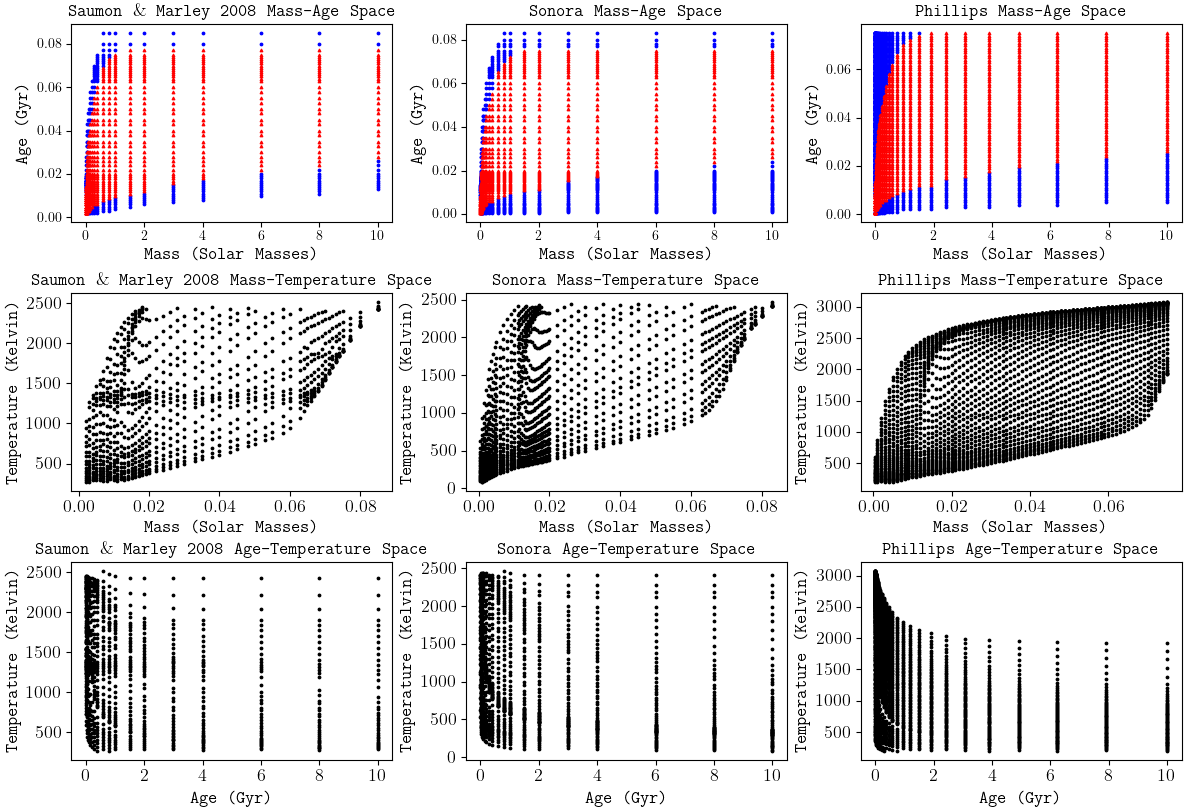}
    \caption{Plots of age vs.\ mass (top row), effective temperature vs.\ mass (middle row), and effective temperature vs.\ age (bottom row) for grid points in the three evolutionary model sets we consider: \cite{saumonmarley2008}, Sonora (\citealt{sonora2021}), and Phillips (\citealt{phillips2020}). The red-colored triangular points in the top row are all evolutionary model points with a temperature between $450$K and $2100$K. In contrast, the top row's circular blue points are those which have temperature values outside of these bounds, namely with temperatures $<450K$ or $>2100K$. 
    \label{fig:evolutionarymodels}}
\end{figure*}

Theoretical models predicting the evolution of brown dwarfs have been formulated, each one presupposing different physics of brown dwarf cooling. In our study we consider the three following evolutionary models: Sonora (\citealt{sonora2021}), Saumon \& Marley 2008 (\citealt{saumonmarley2008}), and Phillips (\citealt{phillips2020}). Figure \ref{fig:evolutionarymodels} shows the grid of cross-sections of the sampled parameter space in mass, age, and temperature covered by each of the evolutionary models.

\par
The particular features of each model relevant to our investigation are delineated below.
\begin{enumerate}
    \item \cite{saumonmarley2008} is our only model that incorporates the effects of dust during the L-T transition, seen as atmospheric cloud cover at the spectral type transition (\citealt{burrows2006}). 
    This model does not include objects that are either massive and young (masses $\geq 0.06 \text{ M}_{\odot}$ and ages $\leq 1$ Gyr), or light and old (masses $\leq 0.0 1\text{ M}_{\odot}$ and ages $\geq 1$ Gyr), as seen by the lack of reference points in the bottom right and top left corners of the top-left subplot in Figure \ref{fig:evolutionarymodels}.
    
    \item The \cite{sonora2021} model features updated chemistry (see their Section 2) but, notably, lacks the earlier assumption of dust and cloud formation during the L-T transition. This model is better sampled than its predecessor for old, light stars, yet it does not extend to objects that are massive and young (mass $\geq 0.06 \text{ M}_{\odot}$ and age $\leq 1$ Gyr).
    
    \item The Phillips (\citealt{phillips2020}) model set offers three evolutionary grids, one using equillibrium chemistry and two using non-equilibrium with differing vertical mixing strenghts, of which we choose to use the evolutionary model with weak mixing. This evolutionary model also does not account for L-T transition dust and cloud formation, although, it is far more thoroughly sampled in the mass-age space than both of the other evolutionary models we consider.

\end{enumerate}

\section{Methods}

\par 
Our primary objective is as follows: determining the best-fit functional form of the substellar mass distribution using the volume-complete sample of brown dwarfs within 20 parsecs of the Sun.

\par
We outline how we create populations with masses and ages consistent with their assumed mass and age distribution (\S 5.1). From there we propagate this population through the evolutionary model (\S 5.2), which provides a present-day value of the effective temperature for each object. All simulations were done in Python, using only fundamental libraries. The source code is available on our Github site\footnote{\url{https://github.com/jgrigorian23/Brown-Dwarf-Simulation-Code}. }.

\subsection{Choosing Mass Functions\label{sec:choosing_mass_functions}}
We choose $\alpha$ values ranging from 0.3 to 0.8 in increments of 0.1 as they envelop the previous best $\alpha$ value of 0.6 (\citealt{kirkpatrick2019}) on either side. Using the $\text{CDF}^{-1}$ of our assumed mass distribution, we pull an object at random from the distribution to assign it a mass. This is done via a Monte Carlo draw from 0 to 1, and we perform $10^6$ of these draws to build a population with statistical robustness. We repeat this procedure for each value of $\alpha$, and for each value of $\alpha$ we repeat the procedure for each of our three assumed low-mass cutoffs. In total, we build eighteen simulated populations, each having masses for $n=10^6$ objects. To be specific, the code samples the mass function again for each different combination of birthrate and evolutionary model, so in total there are 162 simulated populations (Six values of $\alpha\times$Three mass cutoffs$\times$Three birthrates$\times$Three evolutionary models).

\subsection{Constructing Mass-Age Brown Dwarf Populations}
We similarly use the inverse transform method to pull random ages from each of our three assumed age distributions.
This methodology allows for the creation of brown dwarf populations of arbitrary size whose mass distribution will converge to the shape of the transformed function as the sample size approaches a statistically significant value. 

For each population, the $i\textsuperscript{th}$ element in each mass list and the $i\textsuperscript{th}$ element in the age list become the mass and age of the $i\textsuperscript{th}$ object in the simulated population.

\subsection{Deriving Temperatures}
For each object, we wish to find its current-day temperature using our assumed evolutionary model. However, given the discrete sampling of our evolutionary model grids, the simulated values of age and mass for our object are unlikely to have been included directly in the models. We therefore use bilinear interpolation to fill in the sample space between the model's reference points. 
Not all points in the mass-age domain can be mapped using bilinear interpolation. At the edges of the space sampled by each evolutionary model there exist mass-age regions with points that cannot be enclosed within a rectangle of reference points. As shown in the left column of Figure \ref{fig:evolutionarymodels}, each evolutionary model has loci in which we cannot interpolate stellar temperatures.
In such cases we simply disregard the star and assign to the star's temperature value the number $-1$ to indicate that a temperature could not be interpolated. The extent to which we lose objects during the interpolation depends on the (non-)rectilinearity of the provided evolutionary sample set in mass-age space. Both the Sonora and Phillips models are fairly well sampled and do not drop many brown dwarf samples along the whole range of possible mass and age values. In contrast, the \cite{saumonmarley2008} model drops the most objects of our three evolutionary models, especially those sample points which are either young and massive, or old and light, as seen in the top left and bottom right of Figure \ref{fig:evolutionarymodels}. However, it should be noted that for the temperature range we consider for our fitting, namely 450K to 2100K, there are extremely few samples dropped due to a lack of rectilinear bounding reference points (see top row of Figure \ref{fig:evolutionarymodels}) 

Although we state in \S~\ref{sec:choosing_mass_functions}, that we simulated $n = 10^6$ objects, in practice we simulated $n > 10^6$ objects, and kept only the first $10^6$ objects for which temperatures could be obtained via bilinear interpolation.

For our Late-Burst birthrate, we simulate brown dwarf subpopulations as explained at the end of \S 3.3. We shuffle these proportionally sampled constant and Inside-Out birthrate subpopulations and select the first $10^6$ objects to create the Late-Burst birthrate.

\begin{figure}[ht!]
\centering
\includegraphics[scale=0.37]{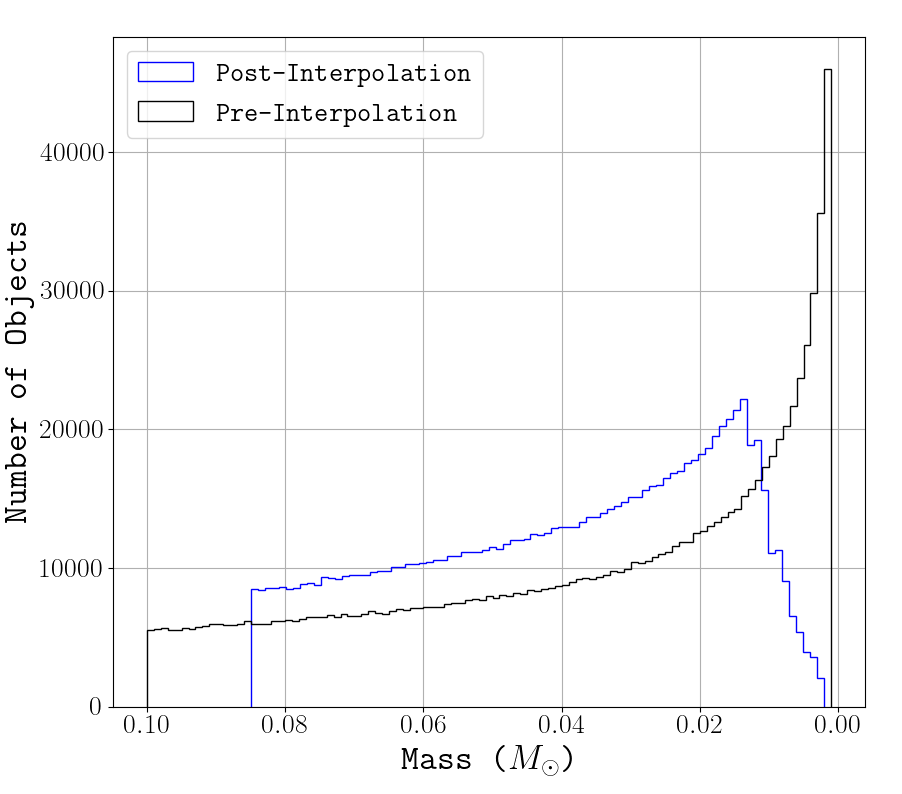}
\caption{The mass distribution of the remaining simulated objects after evolutionary model interpolation for $\alpha=0.5$ with the \cite{saumonmarley2008} evolutionary model and a constant birthrate. The black graph is the original mass distribution with $\alpha=0.5$ and the blue graph is the mass distribution of the remaining objects from the interpolation process. \label{fig:alpha variance masses}}
\end{figure}

The fact that the evolutionary model grids are not sampled over the entire mass and age space needed means that our final, simulated populations contain small biases. See Figure \ref{fig:alpha variance masses} for an example of how our original mass distribution changes after interpolation. 

\section{Findings}

\subsection{Comparison to Empirical Data}

We now shift focus to comparing our simulated populations to the empirical temperature distribution.
We analyze our results in two steps. 
First, in \S 6.1.1 we describe our methods for comparing the simulated and observed temperature distributions to determine which low-mass function $\alpha$ value fits best. Then, in \S 6.1.2, we evaluate which mass cutoff leads to the best fit.

\subsubsection{Temperature Distribution Fitting}

For each simulated brown dwarf population, we consider only those objects with temperatures between $450-2100\text{K}$, as only that range is fully sampled. Many objects in the ranges $300-450\text{K}$ and $2100-2400\text{K}$ are dropped during the interpolation process; i.e., brown dwarfs falling in those ranges are underrepresented because of edges in the model grids.

We obtain our empirical data from \cite{kirkpatrick2023} Table 17, as it provides a volume-limited sample of observed brown dwarfs within 20 parsecs of the Sun. To compare our models against the empirical data, we use the Levenberg-Marquadt algorithm as it is implemented in the IDL routine {\tt mpfit} (\citealt{markwardt2009}). The Levenberg-Marquadt algorithm uniformly scales the simulated population's temperature distribution to find the best fit to the empirical distribution, necessary in  our analysis as our distributions with millions of samples must be appropriately scaled down to be compared against the empirical distribution, which has only a few hundred data points total. 

After many iterations, once the algorithm has optimized the best possible normalization between the two distributions, it returns the minimized residual, quantifying the agreement between the two. We rank our models by their residual value to find the best fit. 

The five best fits for each of the three model suites are listed in Table \ref{table1}. Note that the value of $N$ is the Levenberg-Marquadt normalization constant, unique to each simulated population based on its optimal normalized fitting.

\begin{deluxetable}{cccccc}
\tablenum{1}
\tablecaption{The Five Best Fitting Simulations per Evolutionary Model Set}
\tablewidth{0pt}
\tablehead{
\colhead{Model} & 
\colhead{$\alpha$} & 
\colhead{Birthrate} & 
\colhead{Low-Mass} &  
\colhead{$\chi^2$} &
\colhead{$N$}\\
\colhead{Set\tablenotemark{a}} & 
\colhead{} & 
\colhead{} & 
\colhead{Cutoff} &
\colhead{} &
\colhead{} \\
\colhead{} & 
\colhead{} & 
\colhead{} & 
\colhead{$M_\odot$} &  
\colhead{} &
\colhead{} 
}
\decimalcolnumbers
\startdata
SM08\tablenotemark{b} & 0.5 & Late-Burst & 0.001 & 5.19 & 2427.32  \\
SM08 & 0.6 & Constant & 0.01 & 5.19  & 2362.67\\
SM08 & 0.6 & Constant & 0.001 & 5.22 & 2478.73 \\
SM08 & 0.6 & Constant & 0.005 & 5.24 & 2452.65 \\
SM08 & 0.5 & Constant & 0.001 & 5.38 & 2416.51 \\
\hline 
Sonora & 0.3 & Constant & 0.001 & 21.02 & 2687.96\\
Sonora & 0.3 & Constant & 0.005 & 21.38 & 2422.72\\
Sonora & 0.4 & Constant & 0.001 & 21.44 & 2859.21\\
Sonora & 0.4 & Constant & 0.005 & 21.51 & 2501.77\\
Sonora & 0.3 & Constant & 0.01 & 21.54 & 2183.54\\
\hline
Phillips& 0.3 & Constant & 0.005 & 30.66 & 2270.93 \\
Phillips& 0.4 & Constant & 0.005 & 30.90 & 2357.76\\
Phillips& 0.3 & Constant & 0.001 & 31.01 & 2342.49\\
Phillips& 0.4 & Constant & 0.001 & 31.03 & 2458.24\\
Phillips& 0.5 & Constant & 0.001 & 31.06 & 2591.57\\
\enddata
\label{table1}
\tablenotetext{a}{SM08 = \cite{saumonmarley2008}; Phillips =  \cite{phillips2020}; Sonora = \cite{sonora2021}.}
\tablenotetext{b}{This simulation serves as our choice of the best fitting population.}
\end{deluxetable}

Since the empirical temperature distribution has a bump at the aforementioned L-T transition (1200-1350K), any evolutionary model seeking to be accurate across the gamut of temperature must account for this. Only the \cite{saumonmarley2008} model includes this, so its $\chi^2$ values are naturally the lowest of the three model sets. 

\begin{figure*}
\centering
\includegraphics[scale=0.45]{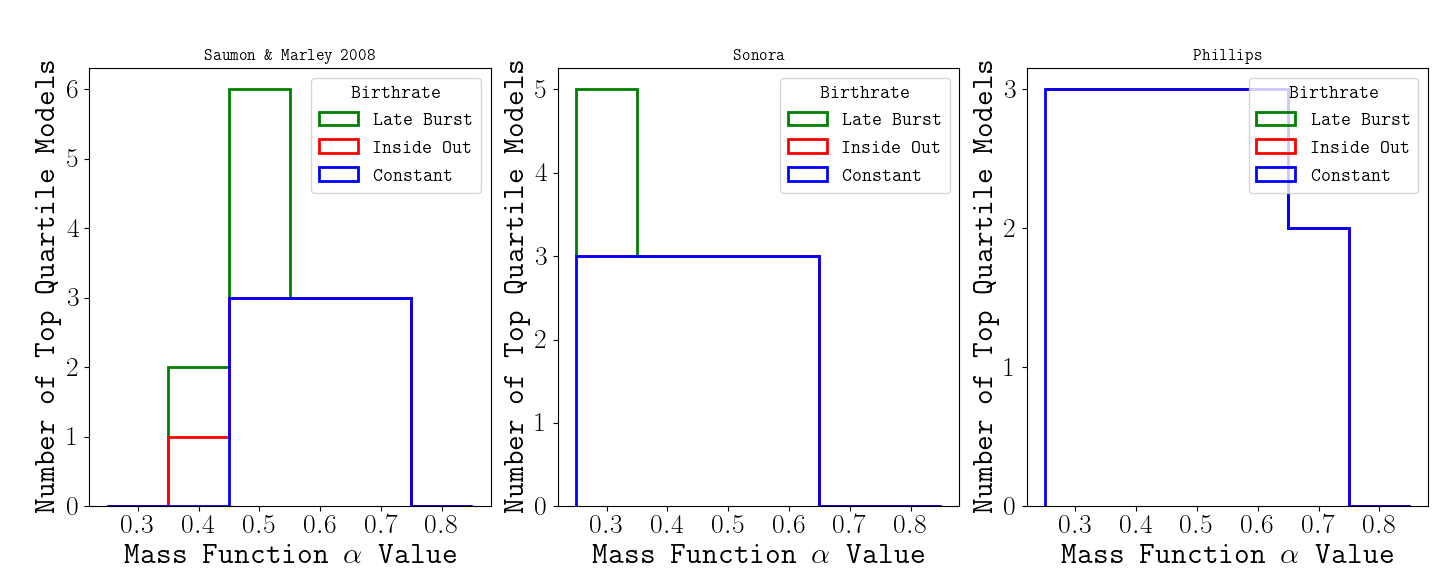}
\caption{The first quartile of best fitting brown dwarf populations colored by their birthrate for each of our evolutionary models.\label{fig:fitting alpha hist}}
\end{figure*}

\begin{figure}[ht!]
\centering
\includegraphics[scale=0.35]{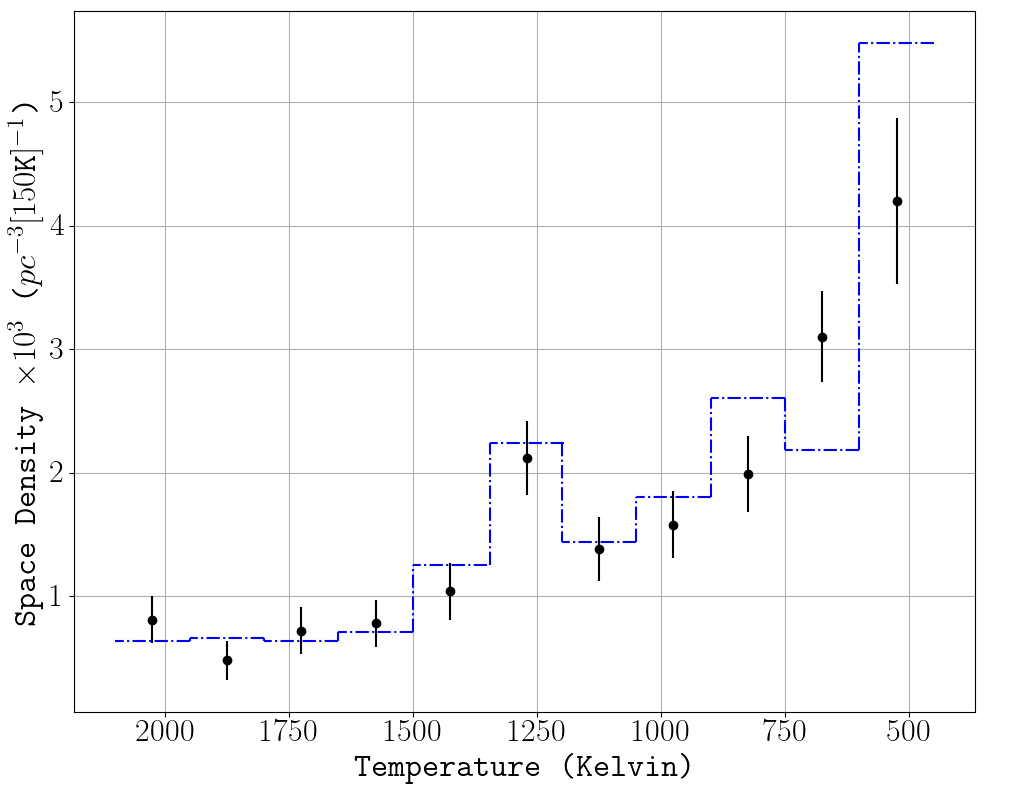}
\caption{Our preferred "best fit" simulation (blue dashed line) -- $\alpha=0.5$, Late-Burst birthrate, low-mass cutoff of $0.001 \text{ M}_{\odot}$, and \cite{saumonmarley2008} evolutionary model grid -- compared to the observed space density of brown dwarfs within the 20-pc census (black points with uncertainties) from \cite{kirkpatrick2023}.
\label{fig:fit example}}
\end{figure}

The $\chi^2$ values of the five best fitting populations displayed in Table \ref{table1} differ only by $0.19$, and thus there are more similarly performing runs not shown in the table that must be considered when constraining the mass function. Figure \ref{fig:fitting alpha hist} shows the $\alpha$ values of the brown dwarf simulations that fall within the first quartile. The conclusion from \cite{kirkpatrick2021} that $\alpha=0.6\pm0.1$ represents the best overall fit was based on a constant birthrate assumption, and as seen by Figure \ref{fig:fitting alpha hist}, this is reproduced in our study, since the distribution of well performing simulations using a constant birthrate is centered around $\alpha=0.6$ as well. Our study, which includes a wider set of birthrates, finds a preferred value of $\alpha=0.5$, based on the peak seen in Figure \ref{fig:fitting alpha hist}. Among the models with $\alpha = 0.5$, the combination that yields the lowest reduced $\chi^2$ is the one given by the \cite{saumonmarley2008} atmospheric models, the Late-Burst birthrate, and a 0.001 $M_\odot$ cutoff. Therefore, we chose this combination as representative of the best overall fit (Figure~\ref{fig:fit example}). We note that \cite{kirkpatrick2023} use a slightly different methodology (see their section 7.1) and find a best fit of $\alpha$ = 0.6 with the constant birthrate.

\subsubsection{Analysis of the Low-Mass Cutoff}
Our second round of analysis focuses on constraining the low-mass cutoff. We move our focus to the cold end of the temperature distribution, as the effects of the cutoff mass are most easily seen here.

The \cite{saumonmarley2008} model grid has very sparse coverage of the lowest masses, so it is not very helpful in determining the low-mass cutoff. However, we can examine the low-mass cutoff using the best fit mass and age distributions along the whole temperature range, $\alpha=0.5$ and the constant birthrate respectively (see left subplot of Figure \ref{fig:fit example}) , along with the best sampled evolutionary model for low masses, the Phillips model, as this allows us to vary the low-mass cutoff specifically to view its impacts. We take the best 5 pairs of $\alpha$ and birthrate from the \cite{saumonmarley2008} Evolutionary Model populations and find the corresponding populations that use the Phillips model instead. The results show that all of the best five $\alpha$ and birthrate models perform best with the $0.001M_{\odot}$ mass cutoff, and three of the five $\alpha$ and birthrate pairs have penultimate best fits with the $0.005M_{\odot}$ mass cutoff. This indicates that the low-mass cutoff is $\lesssim0.005M_{\odot}$

 Our efforts in \S 6.1.1 reveal the combinations of mass function and birthrate that lead to the best fitting temperature distributions. The best performing mass shows a small skew towards the lower mass cutoffs of $0.001 \text{ M}_{\odot}$ and $0.005 \text{ M}_{\odot}$ which correlates with previous result that the low-mass cutoff is at or below $0.005 \text{ M}_{\odot}$ in \cite{kirkpatrick2019}.

\begin{figure}[ht!]
\centering
\includegraphics[scale=0.32]{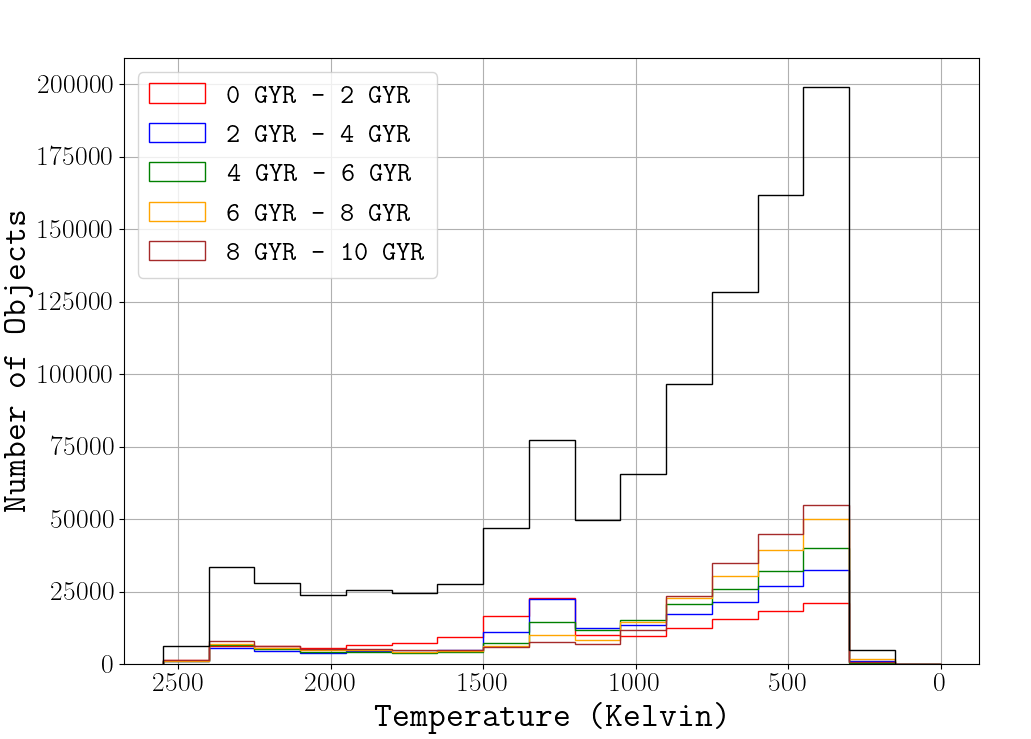}\caption{The temperature distribution of our best fit simulation (black line: $\alpha=0.5$, Late-Burst birthrate, low-mass cutoff of $0.001 \text{ M}_{\odot}$, and \citealt{saumonmarley2008} evolutionary model) decomposed into age regimes (colored lines). \label{fig:temperature decomp}}
\end{figure}

\subsection{L/T Transition\label{sec:LT_transition}}
A key feature of the \cite{saumonmarley2008} evolutionary model is its incorporation of cloud formation at the L-T transition, in which L dwarfs cool into T dwarfs. This process is mainly limited to the temperature range from 1200-1350K, and in this temperature bin there is a noted increase of objects, forming a bump in the temperature distribution (Figure \ref{fig:fit example} and Figure 13 of \citealt{kirkpatrick2019}).
The exact physical conditions and processes that lead to this surplus of objects at 1200-1350K are not yet fully understood, but one theory suggests that the dispersion of the cloud layers could be driven by a radiative cloud-induced variability (\citealt{tanshowman2019}). 

\begin{figure}[ht!]
\centering
\includegraphics[scale=0.35]{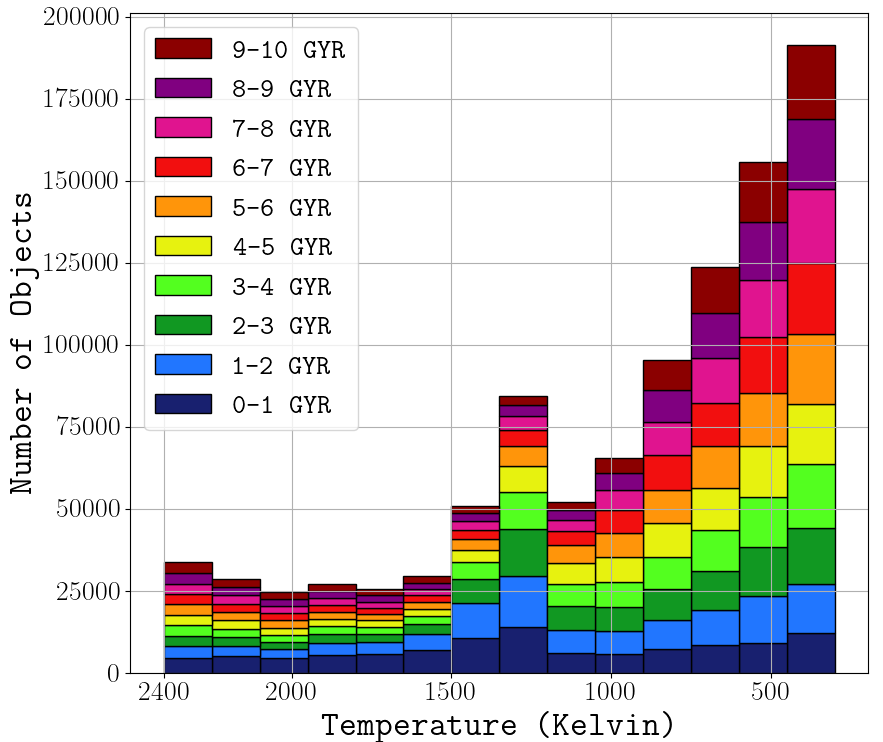}
\caption{The temperature distribution of our best fit simulation ($\alpha=0.5$, Late-Burst birthrate,  low-mass cutoff of $0.001 \text{ M}_{\odot}$, and the \citealt{saumonmarley2008} evolutionary models) color coded by the age of each object. \label{fig:age stack}}
\end{figure}

\begin{figure}[ht!]
\centering
\includegraphics[scale=0.35]{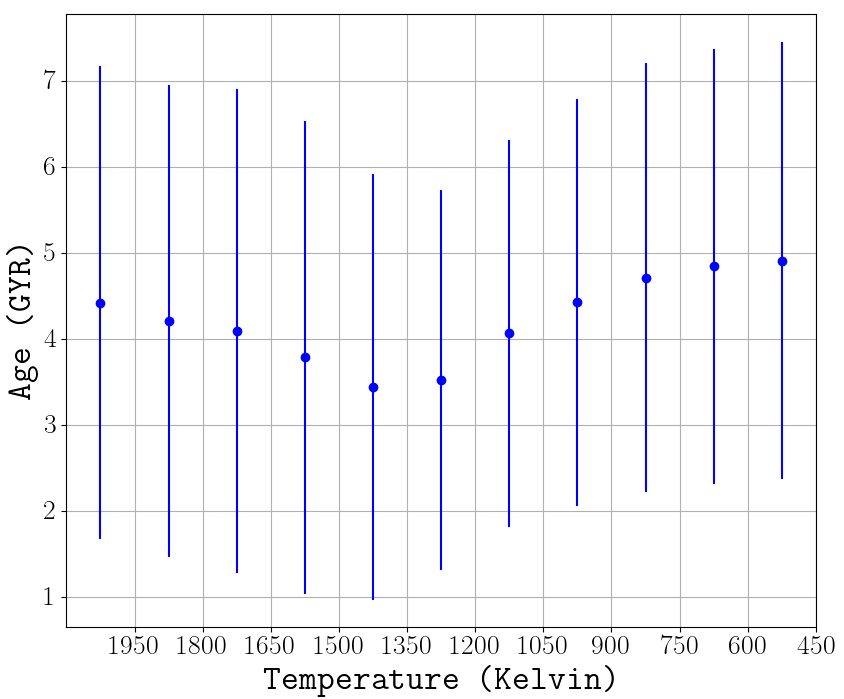}
\caption{The median age and standard deviation in each temperature bin from Figure \ref{fig:age stack}. \label{fig:stdevs}}
\end{figure}

Figures \ref{fig:temperature decomp} and \ref{fig:age stack} show the temperature distribution of our best-fit model colored by object age, showing an excess of young brown dwarfs at the L/T transition bin (1200-1350K). This trend of younger brown dwarfs around the L/T transition is also visible in Figure \ref{fig:stdevs}, where we display the median age and its standard deviation per bin. These predictions suggest a pile-up of young objects just prior to the L/T transition, indicating that the cooling time of a brown dwarf is significantly slowed in this region. Observational confirmation of this effect may be possible once we are able to collect a large field sample of brown dwarf age estimates or, perhaps more easily, measuring the temperature distribution of L and T dwarfs belonging to young clusters and associations of known age.

\subsection{Impacts of Changing $\alpha$, Birthrate, Cutoff, or Model}
The composition of each of our simulated populations depends heavily on our choice of mass function, birthrate, low-mass cutoff, and evolutionary model. In this section, we show the variation in the resulting temperature distribution when we hold all but one of these parameters constant.

The greatest change in the temperature distribution results from a change in the evolutionary model. Notably, simulations from the \cite{saumonmarley2008} models possess a bump at the L/T transition, whereas those from both the Sonora and Phillips models do not.

\begin{figure}[ht!]
\centering
\includegraphics[scale=0.35]{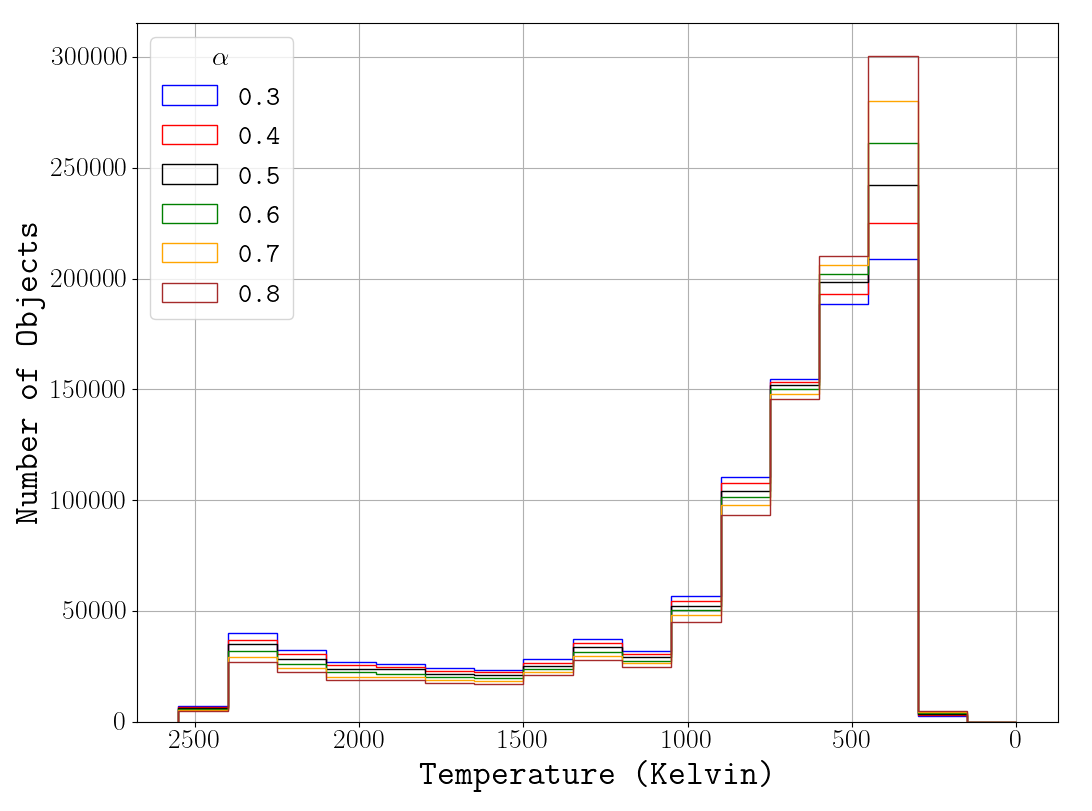}
\caption{Temperature distributions for simulated brown dwarf populations with a varying mass function $\alpha$ parameter ($\alpha\in\{0.3,0.4,0.5,0.6,0.8\}$).  The birthrate is the constant birthrate with a low-mass cutoff of $0.001M_{\odot}$ and \cite{saumonmarley2008} Evolutionary Model. \label{fig:alpha variance temps}}
\end{figure}

As we vary the mass function $\alpha$ parameter, the shape of the temperature distribution also predictably varies (Figure \ref{fig:alpha variance temps}). Flatter mass functions ($\alpha\sim0.4$) lead to temperature distributions with relatively hotter objects whereas steeper mass functions ($\alpha\sim0.7$) lead to a greater abundance of cooler objects. Also, flatter mass functions imply a larger concentration of objects at the L/T transition with a lesser low-temperature peak (300K-600K) and vice versa for steeper mass functions. It should be noted that other differences are marginal everywhere except the low-temperature peak, where the difference between the $\alpha=0.3$ and $\alpha=0.8$ distributions is pronounced. Fundamentally, increasing the mass function's steepness serves to skew the resulting temperature distribution towards the cooler end of the temperature regime.

\begin{figure}[ht!]
\centering
\includegraphics[scale=0.3]{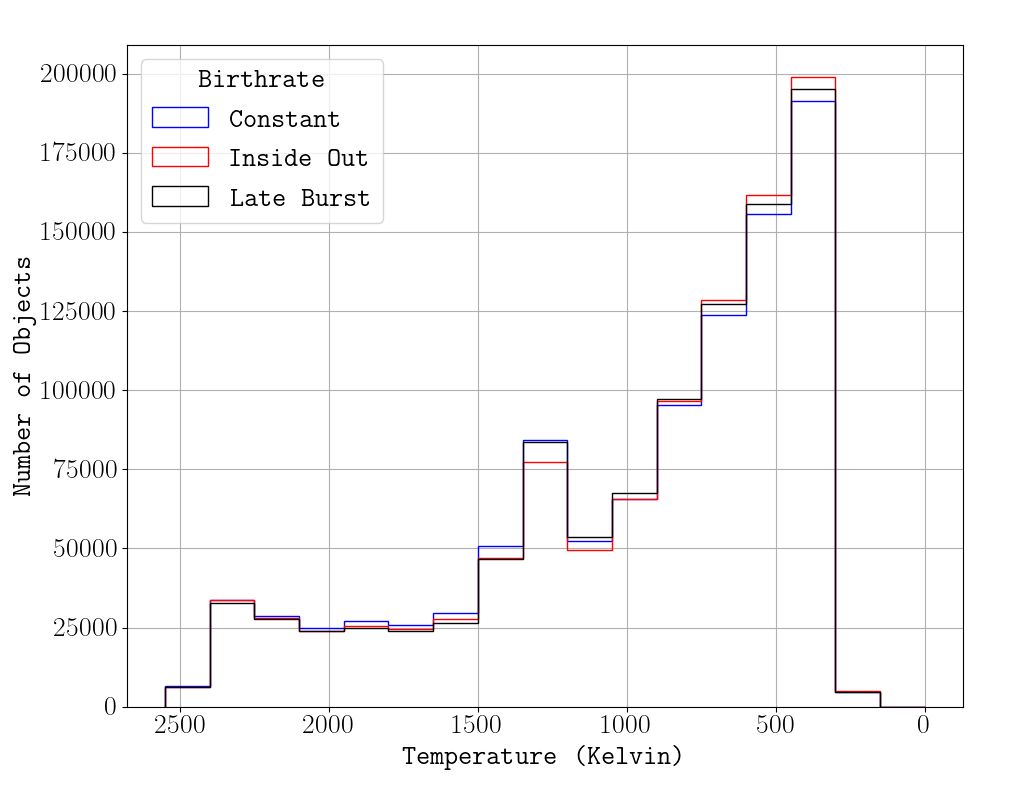}
\caption{Temperature distributions for simulated brown dwarf populations with a varying age function parameter. The low-mass cutoff is 0.001 $M_{\odot}$ with an $\alpha$ of 0.5 and the \cite{saumonmarley2008} Evolutionary Model. The three assumed birthrates are constant (blue), Inside-Out (red), or Late-Burst (black) from \S \ref{sec:age_distributions}.\label{fig:age variance temps}}
\end{figure}

Differences in birthrate functions have already been shown to affect the resulting temperature distribution only marginally (\citealt{burgasser2004}). Our findings corroborate this (Figure \ref{fig:age variance temps}). 

Nonetheless, the Inside-Out age function, for example, allows for a flatter mass function to fit the empirical temperature distribution. The  $\alpha=0.6$ value taken as the ideal mass function steepness in \cite{kirkpatrick2019} assumed a constant birthrate, and our findings in Table \ref{table1} replicate that while also showing that a combination of an $\alpha=0.4$ or $\alpha=0.5$ paired with an Inside-Out birthrate also fit the empirical data quite well. This is because the declining birthrate represented by the Inside-Out function paired with a less steep (lower $\alpha$) mass function can create as many present-day late-T and Y dwarfs as a constant birthrate paired with a steeper mass function.

\begin{figure}[ht!]
\centering
\includegraphics[scale=0.3]{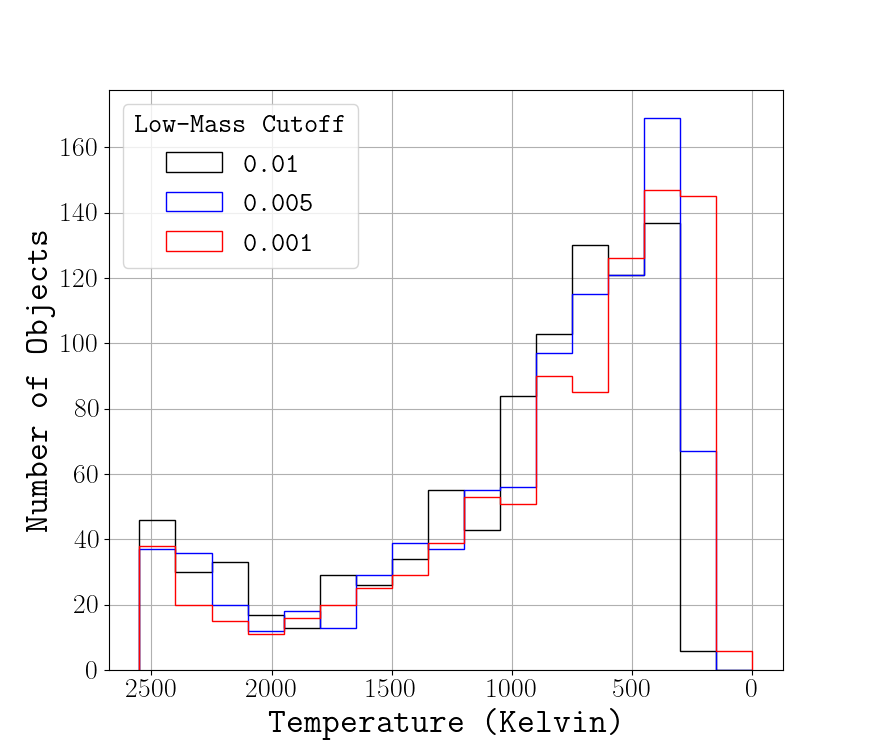}
\caption{Temperature distributions for simulated brown dwarf populations with a varying low-mass cutoff parameter (mass cutoffs:     $0.01\text{ M}_{\odot},0.005\text{ M}_{\odot},0.001\text{ M}_{\odot},$). Mass Function $\alpha$: 0.5, Birthrate: Inside-Out, Evolutionary Model: Phillips \cite{phillips2020}.\label{fig:cut variance temps}}
\end{figure}

The mass cutoff also does not in any significant way affect the shape of the simulated temperature distribution except at the coldest temperatures (Figure \ref{fig:cut variance temps}).

\section{Conclusions}
Our study presents an updated approach to determine the mass function through brown dwarf population simulations. We pose several power law mass functions and combine them with three sample birthrates to create a suite of simulated brown dwarf populations whose temperature distributions we compare to the empirical temperature distribution from \cite{kirkpatrick2021}. Our results indicate a best fit of $\alpha=0.5$ for a birthrate from \cite{johnson2021} (their so-called "Inside-Out" function) that has been steadily declining over the lifetime of the Milky Way, or $\alpha=0.6$ for a constant birthrate, which agrees with a previous study done using the same methodology (\citealt{kirkpatrick2019}).
Our study finds that the low-mass cutoff is $\lesssim0.005M_{\odot}$ by examining the best-performing mass cutoffs. However, tighter error bars on the space density of Y dwarfs within 20 parsecs would place tighter constraints on alpha while also increasing confidence in the low-mass cutoff  and, if a plethora of even colder Y dwarfs is found, push the cutoff value even lower.

All of the code we used to simulate our populations was written in Python and is publicly available on Zenodo\footnote{\url{https://zenodo.org/doi/10.5281/zenodo.11479693}. } (\citealt{code}). Our formalism allows for brown dwarf  population simulations for a given mass function, age function, evolutionary model, and low-mass cutoff. One such use case of this code is outlined in Appendix A, where we modify our birthrate to only include stars with ages 8-10 Gyr. Applications like these are possible because of the flexible nature of our code base as it allows for great customization of the parameters such as mass function and age function. Methodology such as ours is a step towards piecing together the properties of brown dwarfs that are harder to access through direct observation, as one can imagine using evolutionary models with absolute bolometric luminosity measurements instead of empirically derived effective temperatures, as we have done here. Ultimately, it is further observed brown dwarf data that is sorely needed to stimulate more precise theory.

%\facilities{}

%\software{}  

\appendix
\restartappendixnumbering

\section{Present-day Temperature Distribution for Old Brown Dwarfs }
One interesting application of the simulation framework we have built is our ability to tweak the input parameters to explore other physical scenarios. In this section, we examine the predicted present-day temperature distribution of old stars (8-10 Gyr). We choose $\alpha=0.5$, a low-mass cutoff of $0.001 \text{ M}_{\odot}$, and the \cite{saumonmarley2008} evolutionary models. For simplicity, we choose our age function as simply a constant birthrate ranging from 8 to 10 Gyr. With these parameters, we build a temperature distribution via our publicly available code.

\begin{figure*}[h]
%\centering
\gridline{\fig{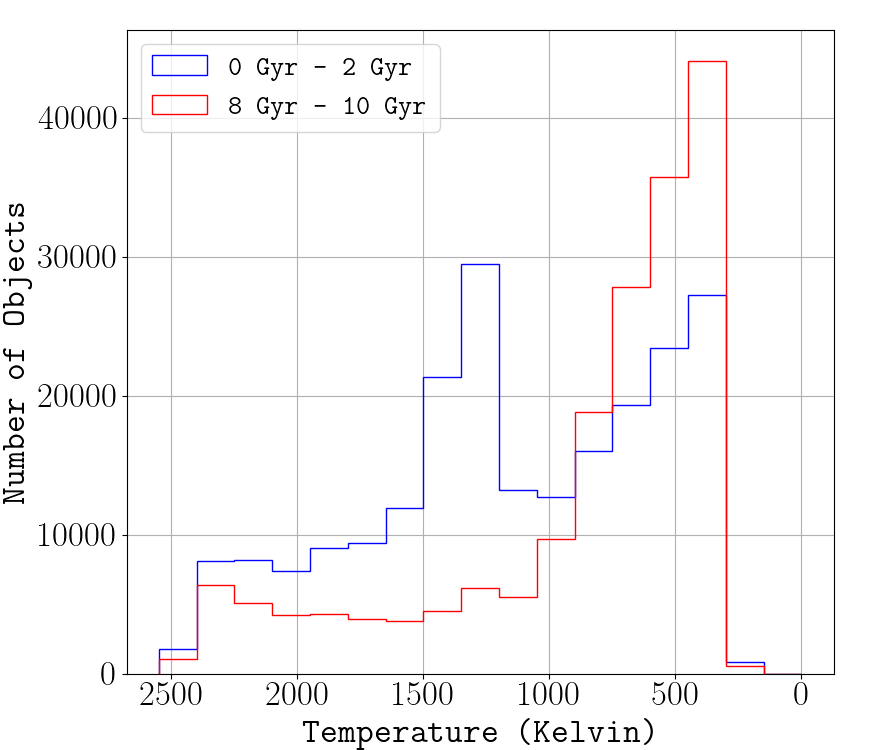}{0.475\textwidth}{(a)}
              \fig{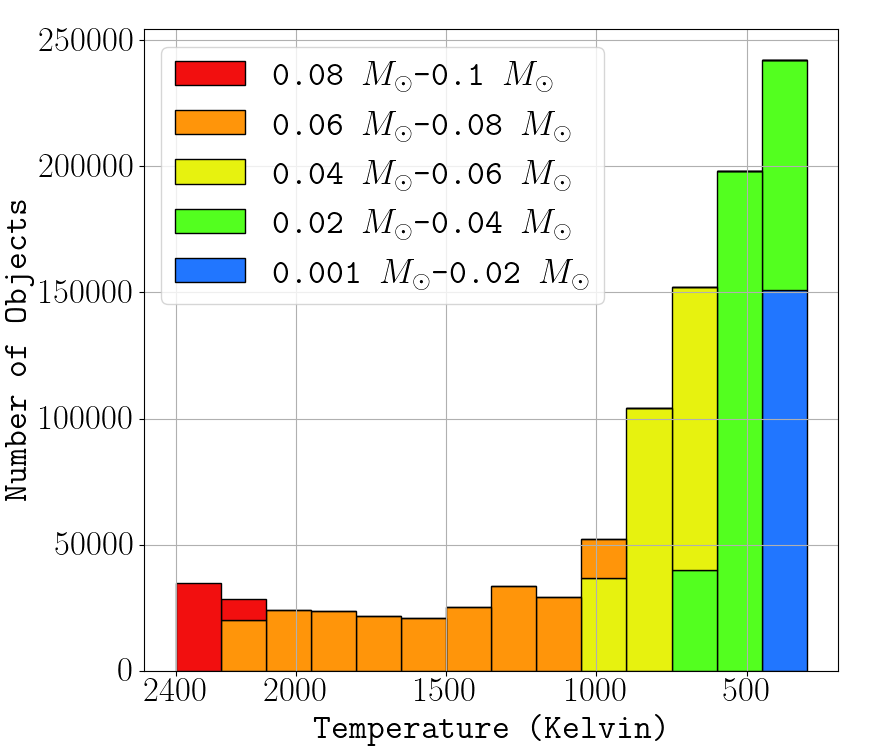}{0.49\textwidth}{(b)}}

\caption{(a) The temperature distribution for young  old brown dwarfs with ages 8-10 Gyr with varying mass cutoffs of $0.01\text{ M}_{\odot}$, $0.005\text{ M}_{\odot}$, or $0.001\text{ M}_{\odot},$) (b) The temperature distribution of objects with ages of 8-10 Gyr further color coded by mass. \label{fig:old hist}}
\end{figure*}

Figure \ref{fig:old hist}a shows this temperature distribution, ultimately revealing how old stars have thermally evolved over time. As discusssed in \S \ref{sec:LT_transition}, the L/T transition bump in the temperature distribution consists mainly of younger objects, so it is no surprise that the temperature distribution of older objects shown here lacks such a bump. Figure \ref{fig:old hist}a shows the change in the temperature distribution for young (0 Gyr - 2 Gyr) brown dwarfs and old (8 Gyr - 10 Gyr) brown dwarfs. Figure \ref{fig:old hist}b agrees with standard knowledge on low-mass brown dwarfs, as they are heavily skewed towards colder temperature bins.

\end{CJK*}

\end{document}